\documentclass[tinyml]{acmart}

\AtBeginDocument{%
  \providecommand\BibTeX{{%
    \normalfont B\kern-0.5em{\scshape i\kern-0.25em b}\kern-0.8em\TeX}}}

\usepackage{diagbox}
\usepackage{multirow}
\usepackage{subcaption}
\usepackage[acronym]{glossaries}
\usepackage{orcidlink}
\usepackage{xcolor}
\usepackage[normalem]{ulem}

\usepackage{siunitx}
\usepackage{icomma}
\usepackage{numprint}

\newcommand{\CRA}[1]{\textcolor{black}{{#1}}}
\newcommand{\CRR}[1]{\textcolor{red}{\sout{}}}

\setcopyright{rightsretained}
\copyrightyear{2024}
\acmYear{2024}

\begin{document}

\title{Boosting keyword spotting through on-device learnable \\ user speech characteristics}

\author{Cristian Cioflan}
\affiliation{%
  \institution{Integrated Systems Laboratory (IIS), \\ ETH Zurich}
  % \city{Zurich}
  % \country{Switzerland}
}
\email{cioflanc@iis.ee.ethz.ch}

\author{Lukas Cavigelli}
\affiliation{%
  \institution{Zurich Research Center, \\ Huawei Technologies}
  % \city{Zurich}
  % \country{Switzerland}
}
\email{lukas.cavigelli@huawei.com}

\author{Luca Benini}
\affiliation{%
  \institution{IIS, ETH Zurich}
  % \city{Zurich}
  % \country{Switzerland}
}
\affiliation{%
  \institution{DEI, University of Bologna}
  % \city{Bologna}
  % \country{Italy}
}
\email{lbenini@iis.ee.ethz.ch}

\begin{abstract}
Keyword spotting systems for always-on TinyML-constrained applications require on-site tuning to boost the accuracy of offline trained classifiers when deployed in unseen inference conditions. Adapting to the speech peculiarities of target users requires many in-domain samples, often unavailable in real-world scenarios. Furthermore, current on-device learning techniques rely on computationally intensive and memory-hungry backbone update schemes, unfit for always-on, battery-powered devices. In this work, we propose a novel on-device learning architecture, composed of a pretrained backbone and a user-aware embedding learning the user's speech characteristics. The so-generated features are fused and used to classify the input utterance. For domain shifts generated by unseen speakers, we measure error rate reductions of up to 19\% from 30.1\% to 24.3\% based on the 35-class problem of the Google Speech Commands dataset, through the inexpensive update of the user projections. We moreover demonstrate the few-shot learning capabilities of our proposed architecture in sample- and class-scarce learning conditions. With \SI{23.7}{\kilo parameters} and 1\,MFLOP per epoch required for on-device training, our system is feasible for TinyML applications aimed at battery-powered microcontrollers. 
\end{abstract}

% Keyword spotting systems for always-on TinyML-constrained applications require on-site tuning to boost the accuracy of offline trained classifiers when deployed in unseen inference conditions. Adapting to the speech peculiarities of target users requires many in-domain samples, often unavailable in real-world scenarios. Furthermore, current on-device learning techniques rely on computationally intensive and memory-hungry backbone update schemes, unfit for always-on, battery-powered devices. In this work, we propose a novel on-device learning architecture, composed of a pretrained backbone and a user-aware embedding learning the user's speech characteristics. The so-generated features are fused and used to classify the input utterance. For domain shifts generated by unseen speakers, we measure error rate reductions of up to 19% from 30.1% to 24.3% based on the 35-class problem of the Google Speech Commands dataset, through the inexpensive update of the user projections. We moreover demonstrate the few-shot learning capabilities of our proposed architecture in sample- and class-scarce learning conditions. With 23.7 kparameters and 1 MFLOP per epoch required for on-device training, our system is feasible for TinyML applications aimed at battery-powered microcontrollers.

\keywords{Keyword Spotting, Embeddings, User Features, On-Device Learning}

\maketitle

\newacronym[plural=FLLs,firstplural=Frequency Locked Loops (FLLs)]{fll}{FLL}{Frequency Locked Loop}
\newacronym{inq}{INQ}{Incremental Network Quantization}
\newacronym{tqt}{TQT}{Training Quantization Thresholds}
\newacronym{ste}{STE}{Straight-Through-Estimator}
\newacronym{bn}{BN}{Batch Normalization}
\newacronym{dma}{DMA}{Direct Memory Access}
\newacronym{simd}{SIMD}{Single Instruction Multiple Data}
\newacronym{lr}{LR}{Learning Rate}
\newacronym[plural=PTUs, firstplural={Pan-Tilt Units}]{ptu}{PTU}{Pan-Tilt Unit}
\newacronym{mdf}{MDF}{Medium-density fibreboard}
\newacronym{soa}{SoA}{State of the Art}
\newacronym{lorawan}{LoRaWAN}{Long Range Wide Area Network}
\newacronym{lora}{LoRa}{Long Range}
\newacronym{dram}{DRAM}{Dynamic Random Access Memory}
\newacronym{fpu}{FPU}{Floating Point Unit}
\newacronym[plural=SCMs, firstplural={Standard Cell Memories (SCMs)}]{scm}{SCM}{Standard Cell Memory}

\newacronym[plural=DVS, firstplural={Dynamic Vision Sensors (DVS)}]{dvs}{DVS}{Dynamic Vision Sensor}
\newacronym[plural=FPGAs, firstplural={Field Programmable Gate Arrays (FPGAs)}]{fpga}{FPGA}{Field Programmable Gate Array}
\newacronym{tpu}{TPU}{Tensor Processing Unit}
\newacronym[plural=WANs, firstplural={Wide Area Networks (WANs)}]{wan}{WAN}{Wide Area Network}
\newacronym[plural=WSNs, firstplural={Wireless Sensor Networks (WSNs)}]{wsn}{WSN}{Wireless Sensor Network}
\newacronym{dl}{DL}{Deep Learning}

\newacronym{fbk}{FBK}{Fondazione Bruno Kessler}
\newacronym{FGSM}{FBK}{Fast Gradient Sign Method}
\newacronym{date}{DATE}{Design Automation and Test in Europe}
\newacronym{iimtc}{I2MTC}{The International Instrumentation \& Measurement Technology Conference}
\newacronym{ini}{INI}{Institute of Neuroinformatics}
\newacronym[plural=LUTs, firstplural={Lookup Tables (LUTs)}]{lut}{LUT}{Lookup Table}

\newacronym{lpwan}{LPWAN}{Low-Power Wide Area Network}
\newacronym{nbiot}{NB-IoT}{Narrow Band Internet-of-Things}

\newacronym{saer}{SAER}{Synchronous Address-Event Representation}
\newacronym{fps}{FPS}{Frames Per Second}
\newacronym{vcd}{VCD}{Value-Change Dump}
\newacronym{spi}{SPI}{Serial Peripheral Interface}
\newacronym{cpi}{CPI}{Camera Parallel Interface}
\newacronym{fifo}{FIFO}{First-In First-Out Queue}
% \newacronym{mcu}{MCU}{Microcontroller}
\newacronym{ble}{BLE}{Bluetooth Low-Energy}
\newacronym{wifi}{Wi-FI}{Wireless Fidelityy}
\newacronym{uart}{UART}{Universal Asynchronous Receiver-Transmitter}
\newacronym{sta}{STA}{Static Timing Analysis}
\newacronym{ptz}{PTZ}{Pan-Tilt Unit}
\newacronym[plural=GPIOs, firstplural={General Purpose Inupt Outputs (GPIOs)}]{gpio}{GPIO}{General Purpose Input Output}
\newacronym[plural=LDOs, firstplural={Low Dropout Regulators (LDOs)}]{ldo}{LDO}{Low Dropout Regulator}

\newacronym{CV}{CV}{Computer Vision}
\newacronym{EoT}{EoT}{Expectation over Transformation}
\newacronym{RPN}{RPN}{Region Proposal Network}
\newacronym{TV}{TV}{Total Variation}
\newacronym{NPS}{NPS}{Non-Printability Score}
\newacronym{STN}{STN}{Spatial Transformer Network}
\newacronym{MTCNN}{MTCNN}{Multi-Task Convolutional Neural Network}
\newacronym{YOLO}{YOLO}{You Only Look Once}
\newacronym{SSD}{SSD}{Single Shot Detector}
\newacronym{SOTA}{SOTA}{State of the Art}
\newacronym{NMS}{NMS}{Non-Maximum Suppression}
\newacronym{ic}{IC}{Integrated Circuit}
\newacronym{rf}{RF}{Radio Frequency}
\newacronym{tcxo}{TCXO}{Temperature Controlled Crystal Oscillator}
\newacronym{jtag}{JTAG}{Joint Test Action Group industry standard}
\newacronym{swd}{SWD}{Serial Wire Debug}
\newacronym{sdio}{SDIO}{Serial Data Input Output}
% \newacronym{ldo}{LDO}{Linear Dropout Regulator}

\newacronym[plural=PCBs, firstplural={Printed Circuit Boards (PCB)}]{pcb}{PCB}{Printed Circuit Board}
\newacronym[plural=ASICs, firstplural={Application Specific Integrated Circuits}]{asic}{ASIC}{Application Specific Integrated Circuit}

\newacronym{ml}{ML}{Machine Learning}
\newacronym{ai}{AI}{Artificial Intelligence}
\newacronym{iot}{IoT}{Internet of Things}
\newacronym{fft}{FFT}{Fast Fourier Transform}
\newacronym[plural=OCUs, firstplural={Output Channel Compute Units (OCUs)}]{ocu}{OCU}{Output Channel Compute Unit}
\newacronym{alu}{ALU}{Arithmetic Logic Unit}
\newacronym{mac}{MAC}{Multiply-Accumulate}
\newacronym{soc}{SoC}{System-on-Chip}

\newacronym{PGD}{PGD}{Projected Gradient Descend}
\newacronym{CW}{CW}{Carlini-Wagner}
\newacronym{OD}{OD}{Object Detection}

\newacronym{rrf}{RRF}{RADAR Repetition Frequency}
\newacronym{nlp}{NLP}{Natural Language Processing}
\newacronym{qam}{QAM}{Quadrature Amplitude Modulation}
\newacronym{rri}{RRI}{RADAR Repetition Interval}
\newacronym{radar}{RADAR}{Radio Detection and Ranging}
\newacronym{loocv}{LOOCV}{Leave-one-out cross validation}
\newacronym{raw}{RAW}{Read-After-Write}
\newacronym[plural=ISAs, firstplural={Instruction Set Architectures (ISAs)}]{isa}{ISA}{Instruction Set Architecture}

\newacronym{os}{OS}{Operating System}
\newacronym{bsp}{BSP}{Board Support Package}
\newacronym{ttn}{TTN}{The Things Network}
\newacronym{wip}{WIP}{Work in Progress}
\newacronym{json}{JSON}{JavaScript Object Notation}
\newacronym{qat}{QAT}{Quantization-Aware Training}

\newacronym{cls}{CLS}{Classification Error}
\newacronym{loc}{LOC}{Localization Error}
\newacronym{bkgd}{BKGD}{Background Error}

\newacronym{dsp}{DSP}{Digital Signal Processing}
\newacronym{mcu}{MCU}{Microcontroller Unit}

\newacronym{gsc}{GSC}{Google Speech Commands}
\newacronym{mswc}{MSWC}{Multilingual Spoken Words Corpus}
\newacronym{demand}{DEMAND}{Diverse Environments Multichannel Acoustic Noise Database}
\newacronym{coco}{COCO}{Common Objects in Context}

\newacronym{asr}{ASR}{Automated Speech Recognition}
\newacronym{kws}{KWS}{Keyword Spotting}
\newacronym{nl-kws}{NL-KWS}{Noiseless Keyword Spotting}
\newacronym{na-kws}{NA-KWS}{Noise-Aware Keyword Spotting}
\newacronym{odda}{ODDA}{On-Device Domain Adaptation}

\newacronym{snr}{SNR}{Signal-to-Noise Ratio}
\newacronym{roc}{ROC}{Receiver Operating Characteristic}
\newacronym{frr}{FRR}{False Rejection Rate}
\newacronym{eer}{EER}{Equal Error Rate}
\newacronym{ce}{CE}{Cross-Entropy}

\newacronym[plural=BNNs, firstplural={Binary Neural Networks (BNNs)}]{bnn}{BNN}{Binary Neural Network}
\newacronym[plural=NNs, firstplural={Neural Networks}]{nn}{NN}{Neural Network (NNs)}
\newacronym[plural=SNNs, firstplural={Spiking Neural Networks (SNNs)}]{snn}{SNN}{Spiking Neural Network}
\newacronym[plural=DNNs, firstplural={Deep Neural Networks (DNNs)}]{dnn}{DNN}{Deep Neural Network}
\newacronym[plural=TCNs,firstplural=Temporal Convolutional Networks]{tcn}{TCN}{Temporal Convolutional Network}
\newacronym[plural=CNNs,firstplural=Convolutional Neural Networks (CNNs)]{cnn}{CNN}{Convolutional Neural Network}
\newacronym[plural=TNNs,firstplural=Ternarized Neural Networks]{tnn}{TNN}{Ternarized Neural Network}
\newacronym{ann}{ANN}{Artificial Neural Networks}
\newacronym{dscnn}{DS-CNN}{Depthwise Separable Convolutional Neural Network}
\newacronym{rnn}{RNN}{Recurrent Neural Network}
\newacronym{gcn}{GCN}{Graph Convolutional Network}
\newacronym{fc}{FC}{Fully Connected}
\newacronym{crnn}{CRNN}{Convolutional Recurrent Neural Network}
\newacronym{mhsa}{MHSA}{Multi-Head Self-attention}
\newacronym{clca}{CLCA}{Convolutional Linear Cross-Attention}
\newacronym{cvat}{CVAT}{Computer Vision Annotation Tool}

\newacronym{mfcc}{MFCC}{Mel-Frequency Cepstral Coefficient}
\newacronym{bf}{BF}{Beamforming}
\newacronym{anc}{ANC}{Active Noise Cancellation}
\newacronym{agc}{AGC}{Automatic Gain Control}
\newacronym{se}{SE}{Speech Enhancement}
\newacronym{mct}{MCT}{Multi-Condition Training}
\newacronym{mcta}{MCTA}{Multi-Condition Training \& Adaptation}
\newacronym{pcen}{PCEN}{Per-Channel Energy Normalization}
\newacronym{sf}{SF}{Sensor Fusion}

\newacronym{mac}{MAC}{Multiply-Accumulate}
\newacronym{flop}{FLOP}{Floating-Point Operation}
\newacronym{fp}{FP}{Floating-Point}

\newacronym{fscil}{FSCIL}{Few-Shot Class-Incremental Learning}
\newacronym{ofscil}{O-FSCIL}{Orthogonal Few-Shot Class-Incremental Learning}
\newacronym{ncfscil}{NC-FSCIL}{Neural Collapse Few-Shot Class-Incremental Learning}
\newacronym{cfscil}{C-FSCIL}{Constrained Few-Shot Class-Incremental Learning}
\newacronym{fsl}{FSL}{Few-Shot Learning}
\newacronym{cil}{CIL}{Class-Incremental Learning}
\newacronym{dil}{DIL}{Domain-Incremental Learning}
\newacronym{savc}{SAVC}{Semantic-Aware Virtual Contrastive}

\newacronym{am}{AM}{Activation Memory}
\newacronym{em}{EM}{Episodic Memory}
\newacronym{fcr}{FCR}{Fully Connected Reductor}
\newacronym{fcc}{FCC}{Fully Connected Classifier}

\newacronym{ol}{OL}{Online Learning}
\newacronym{odl}{ODL}{On_Device Learning}
\section{Introduction}
\label{sec:introduction}

In the last years, the traditional~\gls{ml} paradigm "train-once-deploy-everywhere" started to be challenged~\cite{lin2022ondevice,cioflan2022towards}, as intelligent systems ought to expand their knowledge domain when exposed to novel, dynamic environments. 
Perturbation sources for audio systems are background noises, reverberation, or echo~\cite{indenbom2023deepvqe}, but also include inconspicuous speech characteristics. 
Vocal features (e.g., pitch, intonation)~\cite{guyer2021paralinguistic} or speech disorders (e.g., stuttering) are intrinsic differences between speakers. 
Biases in dataset acquisition, such as gender~\cite{mazumder2021multilingual} or accent~\cite{warden2018speech}, may also reflect in the accuracy when underrepresented speakers are utilizing audio systems. 
Instead of collecting larger, diverse datasets to pretrain larger, generic models, it is more efficient to adapt an already-deployed network to the particular speech properties of the target user.

On-device training frameworks~\cite{lin2022ondevice, ren2021tinyol, nadalini2023reduced} have recently been addressing the need to refine a pretrained model on the target device. 
Often deployed at the extreme edge where available resources are limited, the adaptation must abide by the TinyML constraints (i.e., storage, memory, latency or \glspl{flop}, power consumption)~\cite{banbury2020benchmarking}. 
Classical memory-hungry, computationally expensive backpropagation learning and its variants are unsuitable in such scenarios. 
Furthermore, such strategies often require many labeled samples for reliable fine-tuning, which are unavailable in real-world scenarios.

~\gls{kws} has become increasingly popular in the consumer electronics market, making it convenient for users to interact with their devices using natural language and voice commands.
It is widely used in smart speakers, smartphones, and other consumer~\gls{iot} devices~\cite{hoy2018alexa}.
This work proposes an on-device learning approach for~\gls{kws}, combining a lightweight, frozen backbone with user embeddings. 
Our system can learn particular speech characteristics of the target user, showing improvements in classification error by up to 18\% on out-of-domain speakers. 
We study the impact of the number of samples per class and classes per speaker available during the online learning stage and demonstrate that effective learning can be achieved with as few as four labeled inputs per class. 
We demonstrate that our learning architecture is suitable for ultra-low-power, memory-constrained platforms, requiring storing as few as \SI{4}{\kilo \byte} of intermediate results to update the user features.
\section{Related Work}
\label{sec:relatedwork}

\begin{figure*}[h!]
    \centering
    \includegraphics[width=\linewidth]{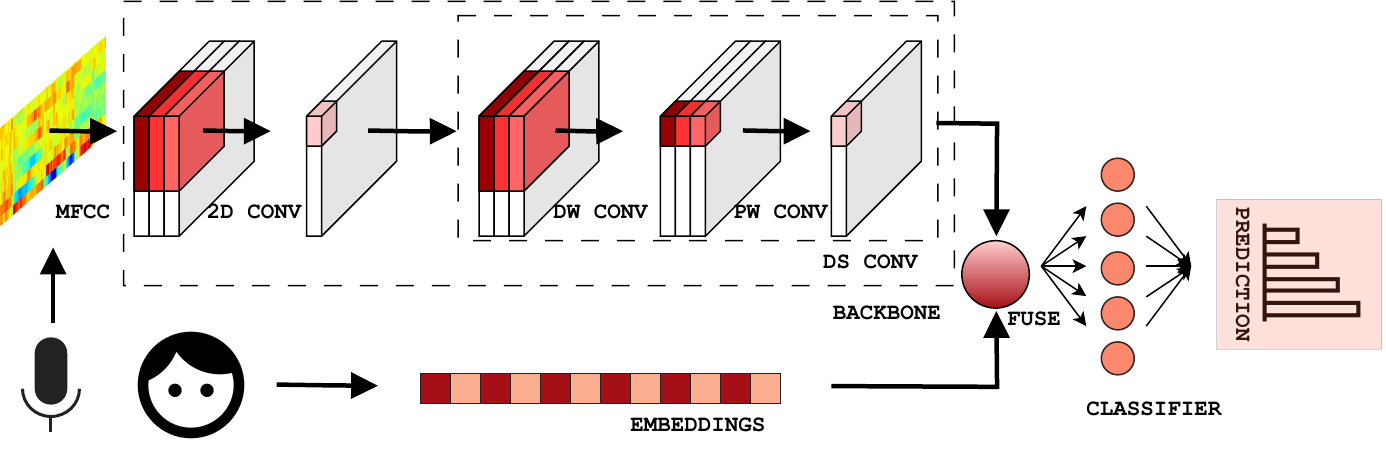}
   
    \caption{\textbf{Architecture overview.} The backbone uses the audio recording to produce activations, whereas the embedding layer uses the unique ID of the speaker to learn user features. The activations and the user features are fused and employed by a fully connected classifier to classify the input utterance.}
    \label{fig:architecture}
\end{figure*}

\subsection{On-Device Keyword Spotting}
\label{ssc:keywordspotting}

Keyword spotting, also closed-vocabulary speech recognition, represents the task of recognizing a set of predefined keywords within a stream of user utterances.
~\gls{kws} has been successfully employed in always-on, voice-activated virtual assistants~\cite{hoy2018alexa} or sound source localization~\cite{manamperi2022drone}. 
Various topologies have been proposed to solve this task, including~\glspl{cnn}~\cite{zhang2017hello},~\glspl{rnn}~\cite{de_andrade_neural_2018}, or transformers~\cite{ding_letr_2022}, with remarkable results on established~\gls{kws} datasets~\cite{warden2018speech}. 
Nonetheless, in the context of battery-powered edge devices performing real-time inference, lightweight \glspl{dnn} are required. 
Notably,~\glspl{dscnn} have been proved feasible for both inference~\cite{mittermaier2020small,zhang2017hello} and on-device training~\cite{cioflan2022towards,nadalini2023reduced}.

\subsection{Speaker-adaptive Audio Systems}
\label{ssc:adaptive}

Speaker adaptation in~\gls{asr} end-to-end frameworks has been explored to mitigate the effect of speaker variations. 
Vector embeddings (e.g., i-vectors~\cite{saon2021advancing}, x-vectors~\cite{snyder2018xvectors}, or s-vectors~\cite{mary2022svectors}) generated for the target speakers are appended to the input features, thus obtaining a speaker-adaptive~\gls{asr} model. 
Such vectors are the latent representation of \glspl{dnn} pretrained on speaker recognition tasks~\cite{mary2022svectors,wagner2023speaker}, thus obtaining a two-stage~\gls{asr} system. 
Such strategies are unsuitable for edge devices due to their increased latency, which can jeopardize the system's real-time capabilities.

To minimize the inference latency, alternative solutions propose refining the backbone to the vocal domain of the target speaker, allowing for a one-time adaptation cost. 
Nevertheless, this strategy has several shortcomings: firstly, refining complex~\glspl{dnn} requires memory amounts often unavailable on~\glspl{mcu}~\cite{lin2022ondevice}, while the computational cost associated with backpropagation should be avoided in order to prolong the lifetime of the~\gls{kws} system. 
Moreover, between~\SI{0.9}{\min}~\cite{sim21robust} and~\SI{30}{\min}~\cite{neekhara2022adapting} of labeled user data are used to improve the system's performance, requiring tedious acquisition and labeling sessions.
Furthermore, when refining the backbone, one must ensure the class balance of the presented data, especially in the context of closed-vocabulary \gls{kws}. 
We propose freezing the backbone and learning only the user embeddings to avoid catastrophic forgetting~\cite{sim21robust} in a class-scarce scenario.
To reduce the retraining cost, accounting for the operations required for gradient computations and the memory needed to store the intermediate results, we integrate the embeddings through late feature-level fusion, extracting features from input data using the two branches and fusing them before the classifier.
\section{User feature-enhanced keyword spotting}
\label{sec:methodology}

With the aim of enabling on-device user adaptation for~\gls{kws}, we develop our system around the~\gls{dscnn} backbone, suitable for both low-resource inference and learning~\cite{cioflan2022towards, zhang2017hello}. 
The backbone comprises an initial 2D convolutional block with an asymmetrical $10\times4$ kernel, aggregating information over a 220~ms window. 
\gls{dscnn} block(s) then extract coarse-grained features, with each block comprising one depthwise convolution and one pointwise convolution, each followed by batch normalization and ReLU layers.
The convolutional features are subsequently used by a~\gls{fc} classifier to produce class predictions. 
With the aim of increasing the system's accuracy, we complement the keyword information by capturing and adapting to the speech characteristics of the target user, as shown in Figure~\ref{fig:architecture}.

To minimize the cost of generating user features, we employ an embedding layer -- a look-up table mapping the speaker identity (i.e., ID) to an $n$-dimensional vector. 
This is more efficient than traditional strategies of storing one copy of the pretrained model and multiple updated copies, one per target speaker.
Instead, for each novel target speaker, we directly add a new entry to the weight matrix, its value trainable end-to-end with the backbone or independently of it. 
The length of each dictionary entry is carefully chosen to match the size of the intermediate activations at the fusion point (i.e., represented as FUSE in Figure~\ref{fig:architecture}) in the backbone.

We merge the user features and the backbone's latent representations before the classifier. 
To minimize the computational effort without compromising accuracy, we study in Section~\ref{ssc:userinformation} different fusion operators (i.e., multiplication, addition, concatenation).
With the fusion performed alongside the channel dimension, we discriminate between two concatenation types: first, backbone-compatible (BC) concatenation, where the length of each dictionary entry matches the number of channels in the backbone (e.g., 64 for \gls{dscnn} S~\cite{zhang2017hello}).
In this scenario, it becomes necessary to modify the input dimensions of the classifier.
Second, classifier-compatible (CC) concatenation, where the number of user features and the number of channels add up to the number of input neurons in the classifier (e.g., 64 for \gls{dscnn} S~\cite{zhang2017hello}).
In this setting, we adapt the number of output channels in the last depthwise separable convolutional layer.

To build a robust~\gls{kws} system, we start by jointly pretraining the backbone and the embedding layer on the~\gls{gsc} dataset~\cite{warden2018speech}. 
The user projections are already present for the pretraining users and fused with the output of the backbone. 
Fusing the features along the channel dimension ensures that each channel projection corresponds to a particular speech characteristic. 
The model can then be deployed in real-world environments, ready to adapt to each target speaker during the~\gls{ol} phase. 

In an on-device adaptation setting, we freeze the backbone and the~\gls{fc} layer to avoid expensive weight updates.
For each target speaker, we presume the availability of a minimum of one sample per class in each of the validation and testing sets, with the remaining utterances collected for a speaker used to update the system. 
Using~\gls{ce} loss, we determine the error generated by the speech characteristics of a new speaker and use it to update only the embedding layer, thus modifying its projection (i.e., depicted in Figure~\ref{fig:architecture} with a checkered pattern).
Notably, the structure of the weight matrix would further enable simultaneous learning of multiple speakers, conditioned by the ability to associate each incoming utterance to a uniquely identifiable speaker.

\section{Results}
\label{sec:results}

\begin{table}[t]
% \footnotesize
\begin{center}
\caption{Validation error rate [\%] after pretraining, leaving out users given their number of recorded samples per class (i.e., 3, 6, or 9), for different embedding fusions. BC and CC represent backbone- and classifier-compatible concatenation.}
\label{tab:pretraining}
\begin{tabular}{c | c c c c} 
 \hline
  &  & GSC10 &  & GSC35 \\ [0.5ex] 
 Embeddings  & 3 & 6 & 9 & 6 \\ [0.5ex] 
 \hline

 - & 6.37 & 4.46 & 5.35 & 27.66 \\ 

 Addition & 6.98 & 4.31 & 4.84 & 25.81 \\

 Multiplication & 6.99 & 4.26 & 4.61 & 25.85 \\
 
 BC Concatenation & 6.77 & 4.29 & 4.91 & 26.54 \\

 CC Concatenation & 7.63 & 4.29 & 5.35 & 26.46 \\ [1ex] 
 \hline

\end{tabular}
\end{center}
\end{table}

To measure the decrease in error rate when performing keyword spotting considering previously unseen speakers, we employ the \gls{gsc} dataset~\cite{warden2018speech}. 
To analyze the impact of task difficulty on accuracy, we evaluate our methodology on both the 10-class problem and the 35-class problem. 
On the latter, we consider the same target classes as~\cite{de_andrade_neural_2018,ding_letr_2022}. 
On the former, in order to uniquely associate each sample with a speaker, we eliminate the \textit{silence} class from the 12-class problem~\cite{warden2018speech}. 
We furthermore eliminate the \textit{unknown} class such that we can ensure the balance of samples per word per speaker, thus remaining with ten target classes: \textit{yes}, \textit{no}, \textit{up}, \textit{down}, \textit{left}, \textit{right}, \textit{on}, \textit{off}, \textit{stop}, and \textit{go}.

We pretrain our networks for up to 40 epochs, employing loss-based early stopping with a patience of 10 epochs, using Adam as an optimizer and plateau-based learning rate scheduling starting with a learning rate of $10^{-3}$. 
We use a batch size of 128. 
For~\gls{ol}, we reduce the initial learning rate to $10^{-5}$ and the patience to 5 epochs, whereas the batch size we set to 10. 
For each new user, we start the learning process from the offline pretrained model, thus avoiding catastrophic forgetting in scenarios involving the update of the weights.

\begin{table*}[ht]
\centering
\caption{Error rate [\%] on~\gls{gsc} 10 and~\gls{gsc} 35 considering the number of samples per class and classes per speaker available for training, for different model updates. The pretraining baselines are 5.33\% on~\gls{gsc} 10 and 30.08\% on~\gls{gsc} 35. Our proposed methodology is highlighted.}
\label{tab:finetuning}
\begin{tabular}{ c c c | c c c c} 
 \hline
    \multirow{2}{*}{\# samples}            & \multirow{2}{*}{Problem} & \multirow{2}{*}{\# classes} & Pretrained &  \textbf{Embedding} & Backbone & Full \\
    & & & model &  \textbf{update only} & update & training \\
    \hline
    \multirow{4}{*}{= 4}     & \multirow{2}{*}{GSC10} & 8 & \multirow{2}{*}{5.33} & \textbf{4.87} & 4.80 & 4.87  \\
                             & & 10 & & \textbf{4.60} & 4.20  & 4.20  \\
                             \cline{2-7}
                             & \multirow{3}{*}{GSC35} & 20 & \multirow{3}{*}{30.08} & \textbf{27.00} & 26.87 & 26.80 \\
                             & & 30 & & \textbf{25.61} & 24.67 & 24.94  \\               
                             & & 35 & & \textbf{24.34} & 24.40 & 24.54 \\
                             \cline{2-7}
    \multirow{4}{*}{$\ge 4$}  & \multirow{2}{*}{GSC10} & 8 & \multirow{2}{*}{5.33} & \textbf{4.41} & 3.61 & 3.27 \\
                             & & 10 & & \textbf{4.34} & 2.14 & 2.00 \\
                             \cline{2-7}
                             & \multirow{3}{*}{GSC35} & 20 & \multirow{3}{*}{30.08} & \textbf{26.94} & 26.80 & 26.54 \\
                             & & 30 & & \textbf{25.54} & 25.74 & 25.41 \\                        
                             & & 35 & & \textbf{24.81} & 24.07 & 24.20 \\  [1ex]                            
    \hline

 \hline
\end{tabular}

\end{table*}

\subsection{User information decreases keyword spotting error}
\label{ssc:userinformation}

To determine the contribution of the user information, modeled through embeddings generated using the unique identifier of a user, we compare the accuracy of a backbone-only pretrained network with that of the speaker-aware model. 
We consider four fusion modalities: addition, multiplication, and concatenation, introduced in Section~\ref{sec:methodology}.
The keyword spotting error rate results for~\gls{gsc} 10 and~\gls{gsc} 35 are shown in Table~\ref{tab:pretraining}.

To assess the impact of available data, we vary the number of speakers available during the pretraining stage.
From a total number of \num{2618} speakers in the \gls{gsc} dataset, each with a different number of recordings per class, we consider for the online learning stage only those speakers with a good representation.
Namely, we keep for the~\gls{ol} problem those users with at least three, six, or nine samples per class in the~\gls{gsc} 10 problem --- i.e., \num{362}, \num{6}, or \num{1} speaker, respectively.
As the \gls{ol} speakers would not be available during the pretraining stage in a real-world scenario, we perform pretraining on the remaining \num{2256}, \num{2612}, or \num{2617} speakers, respectively.
Table~\ref{tab:pretraining} shows the validation error rates obtained after splitting the pretraining set following a 90:10 training:validation ratio, averaged over five runs.

In the first scenario, the unavailability of almost a fifth of the speaker significantly impacts the error rate, as high as 6.37\% for a backbone-only model. 
Surprisingly, when comparing the second and the third pretraining scenarios, the classification error rate is larger when excluding the only speaker with at least nine samples per class, yet we attribute this to the training noise (e.g., weight initialization), since the speaker overlap is 99.8\%.
Nonetheless, in both scenarios, adding user information to the keyword spotting model decreases the error rate to 4.26\%, with the multiplicative integration yielding the lowest error.
On the~\gls{gsc} 35 problem, with no speaker meeting the sample per class criterion, we use the pretraining-\gls{ol} split obtained for~\gls{gsc} 10.
Similarly to the other settings, speaker-aware models show lower error rates, while multiplicative embeddings decrease the error by \CRR{18\%}\CRA{6\% to 25.85\%}, moreover yielding the lowest error when averaged over all four aforementioned scenarios.

\subsection{Few-shot user adaptation}
\label{ssc:fewshot}

Given the scarcity of labeled samples in real-world environments, we study the impact of the number of classes per speaker and samples per class available during training, all while keeping the test set fixed. 
% In all scenarios, we assume the availability of precisely one sample per class in the validation and testing sets. 
Following the conclusions drawn in Section~\ref{ssc:userinformation}, we consider only six speakers, each with at least six samples per class for the online learning scenario. 
Out of the existing samples, one is integrated into the test set, one into the validation set, and the remaining ones (i.e., between 4 and 22) are employed for training. 
Each speaker represents an individual learning session, initialized from the offline pretrained backbone.
We perform five differently seeded runs for each session and report the average error rate over the 30 obtained experiments. 

In Table~\ref{tab:finetuning} we show the testing error rate for the~\gls{gsc} 10 and~\gls{gsc} 35 problems. 
First of all, in each setting, the error rates decrease with the number of training samples available, as more information about the speaker is available for the system.
Compared with the pretrained baselines, learning only the user embeddings decreases the error rate in all scenarios, down to 4.34\% for~\gls{gsc}~10 and 24.34\% for~\gls{gsc}~35\CRA{, a reduction of up to 19\%}.
If only a subset of classes is available during the learning stage (e.g., 20 for~\gls{gsc} 35), relative improvements of 10\% are obtained by learning speech characteristics.

In Table~\ref{tab:finetuning} we furthermore compare our proposed methodology with the results obtained by updating only the backbone or by updating the entire system.
\CRR{When the complete system is refined, user features become particularly important when fewer classes than initially pretrained with are available during the \gls{ol} stage, while relaxing the number of training samples per class.}
\CRA{User features become particularly important when fewer classes than initially pretrained with are available during the \gls{ol} stage, thus creating a domain shift.}
\CRR{Updating both the embeddings and the backbone generates average error reductions of 0.13\% for~\gls{gsc} 10 and 0.1\% for~\gls{gsc} 35 over a backbone-only update, as employing user information mitigates the overfitting potential of the backbone.}
\CRA{In this scenario, while further increasing the number of training samples per class, updating both the embeddings and the backbone generates average error reductions of \CRR{4}\CRA{9.4}\% over a backbone-only update when eight out of~\gls{gsc} 10 classes are available for online learning.
Similarly, complete training improves the accuracy by 0.9\% when 20 out of~\gls{gsc} 35 classes are present, as employing user information mitigates the overfitting risk.}

\begin{table}[t]
\begin{center}
\small
\caption{Error rate reduction on~\gls{gsc}~10 over the pretrained baseline and learning cost per epoch for three~\gls{dscnn} models, considering four training samples per class per speaker. The one-time inference cost of the frozen backbone does not contribute to the cost per epoch for the embedding- and classifier-only update. The pretrained baseline achieves an error rate of 5.33\% on GSC 10.}
\label{tab:modelsize}
\begin{tabular}{c c | c c c} 
 \hline
 & &  & DS-CNN model &  \\ [0.5ex]
 & & S  & M & L \\ [0.5ex] 
  \hline
          & Error decrease [\%] & 1.13 & 1.20 & 1.27 \\
Full      & \glspl{flop} [M] & 354  &  \numprint{2064} &  \numprint{6252}  \\ 
training  & Memory [kB] & \numprint{1530} & \numprint{5467} & \numprint{12998}  \\
          & Energy [$\mu$J] & \numprint{4481} & \numprint{26126} & \numprint{79139}  \\ [1ex]
 \hline
 
  & Error decrease [\%] & 0.47 & 0.73 & 0.40 \\
Classifier & \glspl{flop} [M] & 0.07 & 0.20 & 0.33 \\ 
update & Memory [kB] & 8.1 & 21.5 & 34.4 \\ 
 only  & Energy [$\mu$J] & 0.98 & 2.62 & 4.20 \\ [1ex]

 \hline

 & Error decrease [\%] & 0.73 & 0.94 & 0.30 \\
Embedding & \glspl{flop} [M] & 1.04 & 2.80 & 4.50  \\ 
update & Memory [kB] & 3.6 & 9.7 & 15.5 \\ 
 only  & Energy [$\mu$J] & 13.22 & 35.53 & 57.01 \\ [1ex]

\hline
\end{tabular}
\end{center}

\vspace{-0.4cm}

\end{table}

\begin{figure}[h]
    \centering
    \includegraphics[width=.99\linewidth]{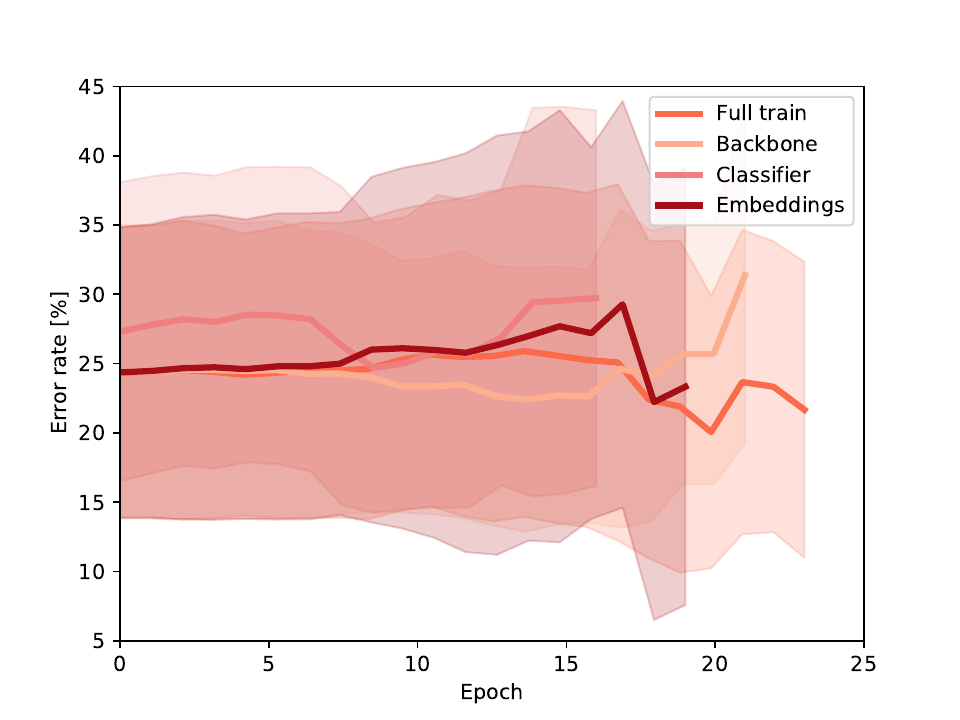}
    \caption{\textbf{Error rate average and standard deviation for~\gls{gsc}~35, comparing learning speech characteristics through embeddings against three update methodologies. \CRA{All methodologies include loss-based early stopping with a patience of 5 epochs.}} }
    \label{fig:learning}
    \vspace{-0.3cm}
\end{figure}

\subsection{Model size and on-device learning}
\label{ssc:modelsize}

To determine the trade-offs between classification error and TinyML-associated constraints, we evaluate our architecture on three incrementally larger backbones, named~\gls{dscnn} S(mall), M(edium), and L(arge)~\cite{zhang2017hello}.
Comprising of \SI{23.7}{}, \SI{138.1}{}, or \SI{416.7}{\kilo parameters} and requiring a number of \SI{2.95}{}, \SI{17.2}{}, or \SI{52.1}{\mega \glspl{flop}} to produce an output, such models are compatible with extreme edge requirements.
In the context of on-device learning, in Table~\ref{tab:modelsize} we indicate the learning cost of fully refining each system through backpropagation, compared with updating only the classifier and with the modest cost of learning the user features.
Given that the backbone is frozen in the latter two scenarios, generating the activations per sample takes place once before the learning process and does not actively contribute to the learning cost.
We additionally \CRR{evaluate}\CRA{estimate} the learning energy cost on the Vega~\cite{rossi2022vega}, a \SI{50}{\milli \watt} specialized SoC with \SI{1.6}{\mega \byte} L2 SRAM and \SI{2}{\mega \byte} of on-chip L3 memory.
Given its power efficiency of \SI{79}{\giga\gls{flop}/\second/\watt}, Vega was shown to be suitable for on-device learning at the extreme edge~\cite{cioflan2022towards}.

Remarkably, solely learning the speech characteristics of the target speaker decreases the classification error for each tested model, the improvements representing up to 78\% of those obtained by fully training the system, all while requiring 340$\times$ fewer~\glspl{flop} per sample.
Furthermore, convergence is achieved faster by only learning the user features instead of refining the entire model, as shown in Figure~\ref{fig:learning}.
The energy cost savings are thus twofold, with \SI{13}{\micro \joule} per epoch required to learn user features through embeddings.
Conversely, although refining the classifier achieves higher energy efficiency than learning user embeddings, the error reduction is 58\% larger by directly learning speech characteristics. 
Therefore, on-device learning of user embeddings represents an optimal accuracy-energy trade-off.  
Through backbone freezing and late feature-level fusion, the (peak) memory requirements do not exceed \SI{16}{\kilo \byte} for our largest model, three orders of magnitude lower than its fully trainable counterpart.
This enables on-device training considering the entire on-chip memory hierarchy, a feature already unattainable for the complete training of \gls{dscnn} M, whose full update requires up to \SI{5.5}{\mega \byte}.

\section{Conclusion}
\label{sec:conclusion}

This work introduces user feature learning for keyword spotting, enabling audio systems to adapt to the speech characteristics of a target speaker. 
Through the multiplicative late fusion of user embeddings into a keyword spotting pipeline, error rate reductions of more than 6 percentage points were measured on both~\gls{gsc} 10 words and 35 words problems. 
Our analyses demonstrate that the proposed learning architecture can increase a system's performance despite not having access to all classes during the online learning stage, while requiring as few as four samples per class. 
With the embeddings' update per sample representing less than 0.3\% of the inference cost and memory requirements under~\SI{4}{\kilo \byte}, our architecture is suitable for battery-powered, memory-constrained platforms, enabling speaker-aware keyword spotting at the extreme edge.

\begin{acks}
\CRA{This work was partly supported by the Swiss National Science Foundation under grant No 207913: TinyTrainer: On-chip Training for TinyML devices.}
\end{acks}

\bibliographystyle{ACM-Reference-Format}
\bibliography{tinyml}


\begin{thebibliography}{22}

%%% the Web address.

\ifx \showCODEN    \undefined \def \showCODEN     #1{\unskip}     \fi
\ifx \showDOI      \undefined \def \showDOI       #1{#1}\fi
\ifx \showISBNx    \undefined \def \showISBNx     #1{\unskip}     \fi
\ifx \showISBNxiii \undefined \def \showISBNxiii  #1{\unskip}     \fi
\ifx \showISSN     \undefined \def \showISSN      #1{\unskip}     \fi
\ifx \showLCCN     \undefined \def \showLCCN      #1{\unskip}     \fi
\ifx \shownote     \undefined \def \shownote      #1{#1}          \fi
\ifx \showarticletitle \undefined \def \showarticletitle #1{#1}   \fi
\ifx \showURL      \undefined \def \showURL       {\relax}        \fi
% invisible to TeX
\providecommand\bibfield[2]{#2}
\providecommand\bibinfo[2]{#2}
\providecommand\natexlab[1]{#1}
\providecommand\showeprint[2][]{arXiv:#2}

\bibitem[Banbury et~al\mbox{.}(2021)]%
        {banbury2020benchmarking}
\bibfield{author}{\bibinfo{person}{Colby~R Banbury},
  \bibinfo{person}{Vijay~Janapa Reddi}, \bibinfo{person}{Max Lam},
  \bibinfo{person}{William Fu}, \bibinfo{person}{Amin Fazel},
  \bibinfo{person}{Jeremy Holleman}, \bibinfo{person}{Xinyuan Huang},
  \bibinfo{person}{Robert Hurtado}, \bibinfo{person}{David Kanter},
  \bibinfo{person}{Anton Lokhmotov}, {et~al\mbox{.}}}
  \bibinfo{year}{2021}\natexlab{}.
\newblock \showarticletitle{Benchmarking tinyml systems: Challenges and
  direction}.
\newblock \bibinfo{journal}{\emph{arXiv}}  \bibinfo{volume}{abs/2003.04821}
  (\bibinfo{year}{2021}).
\newblock


\bibitem[Cioflan et~al\mbox{.}(2022)]%
        {cioflan2022towards}
\bibfield{author}{\bibinfo{person}{Cristian Cioflan}, \bibinfo{person}{Lukas
  Cavigelli}, \bibinfo{person}{Manuele Rusci}, \bibinfo{person}{Miguel
  De~Prado}, {and} \bibinfo{person}{Luca Benini}.}
  \bibinfo{year}{2022}\natexlab{}.
\newblock \showarticletitle{Towards On-device Domain Adaptation for
  Noise-Robust Keyword Spotting}. In \bibinfo{booktitle}{\emph{2022 IEEE 4th
  International Conference on Artificial Intelligence Circuits and Systems
  (AICAS)}}. \bibinfo{pages}{82--85}.
\newblock
\urldef\tempurl%
\url{https://doi.org/10.1109/AICAS54282.2022.9869990}
\showDOI{\tempurl}


\bibitem[de~Andrade et~al\mbox{.}(2018)]%
        {de_andrade_neural_2018}
\bibfield{author}{\bibinfo{person}{Douglas~Coimbra de Andrade},
  \bibinfo{person}{Sabato Leo}, \bibinfo{person}{Martin Loesener Da~Silva
  Viana}, {and} \bibinfo{person}{Christoph Bernkopf}.}
  \bibinfo{year}{2018}\natexlab{}.
\newblock \showarticletitle{A neural attention model for speech command
  recognition}.
\newblock \bibinfo{journal}{\emph{arXiv}}  \bibinfo{volume}{abs/1808.08929}
  (\bibinfo{year}{2018}).
\newblock


\bibitem[Ding et~al\mbox{.}(2022)]%
        {ding_letr_2022}
\bibfield{author}{\bibinfo{person}{Kevin Ding}, \bibinfo{person}{Martin Zong},
  \bibinfo{person}{Jiakui Li}, {and} \bibinfo{person}{Baoxiang Li}.}
  \bibinfo{year}{2022}\natexlab{}.
\newblock \showarticletitle{{LETR}: {A} {Lightweight} and {Efficient}
  {Transformer} for {Keyword} {Spotting}}. In
  \bibinfo{booktitle}{\emph{{ICASSP} 2022 - 2022 {IEEE} {International}
  {Conference} on {Acoustics}, {Speech} and {Signal} {Processing} ({ICASSP})}}.
  \bibinfo{pages}{7987--7991}.
\newblock
\urldef\tempurl%
\url{https://doi.org/10.1109/ICASSP43922.2022.9747295}
\showDOI{\tempurl}
\newblock
\shownote{ISSN: 2379-190X}.


\bibitem[Guyer et~al\mbox{.}(2021)]%
        {guyer2021paralinguistic}
\bibfield{author}{\bibinfo{person}{Joshua~J Guyer}, \bibinfo{person}{Pablo
  Bri{\~n}ol}, \bibinfo{person}{Thomas~I Vaughan-Johnston},
  \bibinfo{person}{Leandre~R Fabrigar}, \bibinfo{person}{Lorena Moreno}, {and}
  \bibinfo{person}{Richard~E Petty}.} \bibinfo{year}{2021}\natexlab{}.
\newblock \showarticletitle{Paralinguistic features communicated through voice
  can affect appraisals of confidence and evaluative judgments}.
\newblock \bibinfo{journal}{\emph{J. Nonverbal Behav.}} \bibinfo{volume}{45},
  \bibinfo{number}{4} (\bibinfo{date}{July} \bibinfo{year}{2021}),
  \bibinfo{pages}{479--504}.
\newblock


\bibitem[Hoy(2018)]%
        {hoy2018alexa}
\bibfield{author}{\bibinfo{person}{Matthew~B. Hoy}.}
  \bibinfo{year}{2018}\natexlab{}.
\newblock \showarticletitle{Alexa, Siri, Cortana, and More: An Introduction to
  Voice Assistants}.
\newblock \bibinfo{journal}{\emph{Medical Reference Services Quarterly}}
  \bibinfo{volume}{37}, \bibinfo{number}{1} (\bibinfo{year}{2018}),
  \bibinfo{pages}{81--88}.
\newblock


\bibitem[Indenbom et~al\mbox{.}(2023)]%
        {indenbom2023deepvqe}
\bibfield{author}{\bibinfo{person}{Evgenii Indenbom}, \bibinfo{person}{Nicolae
  Ristea}, \bibinfo{person}{Ando Saabas}, \bibinfo{person}{Tanel Pärnamaa},
  \bibinfo{person}{Jegor Guzhvin}, {and} \bibinfo{person}{Ross Cutler}.}
  \bibinfo{year}{2023}\natexlab{}.
\newblock \showarticletitle{DeepVQE: Real Time Deep Voice Quality Enhancement
  for Joint Acoustic Echo Cancellation, Noise Suppression and Dereverberation}.
  \bibinfo{pages}{3819--3823}.
\newblock
\urldef\tempurl%
\url{https://doi.org/10.21437/Interspeech.2023-1028}
\showDOI{\tempurl}


\bibitem[Lin et~al\mbox{.}(2022)]%
        {lin2022ondevice}
\bibfield{author}{\bibinfo{person}{Ji Lin}, \bibinfo{person}{Ligeng Zhu},
  \bibinfo{person}{Wei-Ming Chen}, \bibinfo{person}{Wei-Chen Wang},
  \bibinfo{person}{Chuang Gan}, {and} \bibinfo{person}{Song Han}.}
  \bibinfo{year}{2022}\natexlab{}.
\newblock \showarticletitle{On-Device Training Under 256KB Memory}. In
  \bibinfo{booktitle}{\emph{Annual Conference on Neural Information Processing
  Systems (NeurIPS)}}.
\newblock


\bibitem[Manamperi et~al\mbox{.}(2022)]%
        {manamperi2022drone}
\bibfield{author}{\bibinfo{person}{Wageesha Manamperi},
  \bibinfo{person}{Thushara~D. Abhayapala}, \bibinfo{person}{Jihui Zhang},
  {and} \bibinfo{person}{Prasanga~N. Samarasinghe}.}
  \bibinfo{year}{2022}\natexlab{}.
\newblock \showarticletitle{Drone Audition: Sound Source Localization Using
  On-Board Microphones}.
\newblock \bibinfo{journal}{\emph{IEEE/ACM Transactions on Audio, Speech, and
  Language Processing}}  \bibinfo{volume}{30} (\bibinfo{year}{2022}),
  \bibinfo{pages}{508--519}.
\newblock
\urldef\tempurl%
\url{https://doi.org/10.1109/TASLP.2022.3140550}
\showDOI{\tempurl}


\bibitem[Mary et~al\mbox{.}(2021)]%
        {mary2022svectors}
\bibfield{author}{\bibinfo{person}{Narla John Metilda~Sagaya Mary},
  \bibinfo{person}{Srinivasan Umesh}, {and} \bibinfo{person}{Sandesh~Varadaraju
  Katta}.} \bibinfo{year}{2021}\natexlab{}.
\newblock \showarticletitle{S-Vectors and TESA: Speaker Embeddings and a
  Speaker Authenticator Based on Transformer Encoder}.
\newblock \bibinfo{journal}{\emph{IEEE/ACM Trans. Audio, Speech and Lang.
  Proc.}}  \bibinfo{volume}{30} (\bibinfo{date}{dec} \bibinfo{year}{2021}),
  \bibinfo{pages}{404–413}.
\newblock
\showISSN{2329-9290}
\urldef\tempurl%
\url{https://doi.org/10.1109/TASLP.2021.3134566}
\showDOI{\tempurl}


\bibitem[Mazumder et~al\mbox{.}(2021)]%
        {mazumder2021multilingual}
\bibfield{author}{\bibinfo{person}{Mark Mazumder}, \bibinfo{person}{Sharad
  Chitlangia}, \bibinfo{person}{Colby Banbury}, \bibinfo{person}{Yiping Kang},
  \bibinfo{person}{Juan~Manuel Ciro}, \bibinfo{person}{Keith Achorn},
  \bibinfo{person}{Daniel Galvez}, \bibinfo{person}{Mark Sabini},
  \bibinfo{person}{Peter Mattson}, \bibinfo{person}{David Kanter},
  \bibinfo{person}{Greg Diamos}, \bibinfo{person}{Pete Warden},
  \bibinfo{person}{Josh Meyer}, {and} \bibinfo{person}{Vijay~Janapa Reddi}.}
  \bibinfo{year}{2021}\natexlab{}.
\newblock \showarticletitle{Multilingual Spoken Words Corpus}. In
  \bibinfo{booktitle}{\emph{Thirty-fifth Conference on Neural Information
  Processing Systems Datasets and Benchmarks Track (Round 2)}}.
\newblock
\urldef\tempurl%
\url{https://openreview.net/forum?id=c20jiJ5K2H}
\showURL{%
\tempurl}


\bibitem[Mittermaier et~al\mbox{.}(2020)]%
        {mittermaier2020small}
\bibfield{author}{\bibinfo{person}{Simon Mittermaier}, \bibinfo{person}{Ludwig
  Kürzinger}, \bibinfo{person}{Bernd Waschneck}, {and}
  \bibinfo{person}{Gerhard Rigoll}.} \bibinfo{year}{2020}\natexlab{}.
\newblock \showarticletitle{Small-Footprint Keyword Spotting on Raw Audio Data
  with Sinc-Convolutions}. In \bibinfo{booktitle}{\emph{ICASSP 2020 - 2020 IEEE
  International Conference on Acoustics, Speech and Signal Processing
  (ICASSP)}}. \bibinfo{pages}{7454--7458}.
\newblock
\urldef\tempurl%
\url{https://doi.org/10.1109/ICASSP40776.2020.9053395}
\showDOI{\tempurl}


\bibitem[Nadalini et~al\mbox{.}(2023)]%
        {nadalini2023reduced}
\bibfield{author}{\bibinfo{person}{Davide Nadalini}, \bibinfo{person}{Manuele
  Rusci}, \bibinfo{person}{Luca Benini}, {and} \bibinfo{person}{Francesco
  Conti}.} \bibinfo{year}{2023}\natexlab{}.
\newblock \showarticletitle{Reduced precision floating-point optimization for
  Deep Neural Network On-Device Learning on microcontrollers}.
\newblock \bibinfo{journal}{\emph{Future Generation Computer Systems}}
  \bibinfo{volume}{149} (\bibinfo{year}{2023}), \bibinfo{pages}{212--226}.
\newblock
\showISSN{0167-739X}
\urldef\tempurl%
\url{https://doi.org/10.1016/j.future.2023.07.020}
\showDOI{\tempurl}


\bibitem[Neekhara et~al\mbox{.}(2021)]%
        {neekhara2022adapting}
\bibfield{author}{\bibinfo{person}{Paarth Neekhara}, \bibinfo{person}{Jason
  Li}, {and} \bibinfo{person}{Boris Ginsburg}.}
  \bibinfo{year}{2021}\natexlab{}.
\newblock \showarticletitle{Adapting TTS models For New Speakers using Transfer
  Learning}.
\newblock \bibinfo{journal}{\emph{arXiv}}  \bibinfo{volume}{abs/2110.05798}
  (\bibinfo{year}{2021}).
\newblock


\bibitem[Ren et~al\mbox{.}(2021)]%
        {ren2021tinyol}
\bibfield{author}{\bibinfo{person}{Haoyu Ren}, \bibinfo{person}{Darko Anicic},
  {and} \bibinfo{person}{Thomas~A. Runkler}.} \bibinfo{year}{2021}\natexlab{}.
\newblock \showarticletitle{TinyOL: TinyML with Online-Learning on
  Microcontrollers}. In \bibinfo{booktitle}{\emph{2021 International Joint
  Conference on Neural Networks (IJCNN)}}. \bibinfo{pages}{1--8}.
\newblock
\urldef\tempurl%
\url{https://doi.org/10.1109/IJCNN52387.2021.9533927}
\showDOI{\tempurl}


\bibitem[Rossi et~al\mbox{.}(2022)]%
        {rossi2022vega}
\bibfield{author}{\bibinfo{person}{Davide Rossi}, \bibinfo{person}{Francesco
  Conti}, \bibinfo{person}{Manuel Eggiman}, \bibinfo{person}{Alfio~Di Mauro},
  \bibinfo{person}{Giuseppe Tagliavini}, \bibinfo{person}{Stefan Mach},
  \bibinfo{person}{Marco Guermandi}, \bibinfo{person}{Antonio Pullini},
  \bibinfo{person}{Igor Loi}, \bibinfo{person}{Jie Chen}, \bibinfo{person}{Eric
  Flamand}, {and} \bibinfo{person}{Luca Benini}.}
  \bibinfo{year}{2022}\natexlab{}.
\newblock \showarticletitle{Vega: A Ten-Core SoC for IoT Endnodes With DNN
  Acceleration and Cognitive Wake-Up From MRAM-Based State-Retentive Sleep
  Mode}.
\newblock \bibinfo{journal}{\emph{IEEE Journal of Solid-State Circuits}}
  \bibinfo{volume}{57}, \bibinfo{number}{1} (\bibinfo{year}{2022}),
  \bibinfo{pages}{127--139}.
\newblock
\urldef\tempurl%
\url{https://doi.org/10.1109/JSSC.2021.3114881}
\showDOI{\tempurl}


\bibitem[Saon et~al\mbox{.}(2021)]%
        {saon2021advancing}
\bibfield{author}{\bibinfo{person}{George Saon}, \bibinfo{person}{Zoltán
  Tüske}, \bibinfo{person}{Daniel Bolanos}, {and} \bibinfo{person}{Brian
  Kingsbury}.} \bibinfo{year}{2021}\natexlab{}.
\newblock \showarticletitle{Advancing RNN Transducer Technology for Speech
  Recognition}. In \bibinfo{booktitle}{\emph{ICASSP 2021 - 2021 IEEE
  International Conference on Acoustics, Speech and Signal Processing
  (ICASSP)}}. \bibinfo{pages}{5654--5658}.
\newblock
\urldef\tempurl%
\url{https://doi.org/10.1109/ICASSP39728.2021.9414716}
\showDOI{\tempurl}


\bibitem[Sim et~al\mbox{.}(2021)]%
        {sim21robust}
\bibfield{author}{\bibinfo{person}{Khe~Chai Sim}, \bibinfo{person}{Angad
  Chandorkar}, \bibinfo{person}{Fan Gao}, \bibinfo{person}{Mason Chua},
  \bibinfo{person}{Tsendsuren Munkhdalai}, {and} \bibinfo{person}{Françoise
  Beaufays}.} \bibinfo{year}{2021}\natexlab{}.
\newblock \showarticletitle{{Robust Continuous On-Device Personalization for
  Automatic Speech Recognition}}. In \bibinfo{booktitle}{\emph{Proc.
  Interspeech 2021}}. \bibinfo{pages}{1284--1288}.
\newblock
\urldef\tempurl%
\url{https://doi.org/10.21437/Interspeech.2021-318}
\showDOI{\tempurl}


\bibitem[Snyder et~al\mbox{.}(2018)]%
        {snyder2018xvectors}
\bibfield{author}{\bibinfo{person}{David Snyder}, \bibinfo{person}{Daniel
  Garcia-Romero}, \bibinfo{person}{Gregory Sell}, \bibinfo{person}{Daniel
  Povey}, {and} \bibinfo{person}{Sanjeev Khudanpur}.}
  \bibinfo{year}{2018}\natexlab{}.
\newblock \showarticletitle{X-Vectors: Robust DNN Embeddings for Speaker
  Recognition}. In \bibinfo{booktitle}{\emph{2018 IEEE International Conference
  on Acoustics, Speech and Signal Processing (ICASSP)}}.
  \bibinfo{pages}{5329--5333}.
\newblock
\urldef\tempurl%
\url{https://doi.org/10.1109/ICASSP.2018.8461375}
\showDOI{\tempurl}


\bibitem[Wagner et~al\mbox{.}(2023)]%
        {wagner2023speaker}
\bibfield{author}{\bibinfo{person}{Dominik Wagner}, \bibinfo{person}{Ilja
  Baumann}, \bibinfo{person}{Sebastian~P. Bayerl}, \bibinfo{person}{Korbinian
  Riedhammer}, {and} \bibinfo{person}{Tobias Bocklet}.}
  \bibinfo{year}{2023}\natexlab{}.
\newblock \showarticletitle{Speaker Adaptation for End-to-End Speech
  Recognition Systems in Noisy Environments}. In \bibinfo{booktitle}{\emph{2023
  IEEE Automatic Speech Recognition and Understanding Workshop (ASRU)}}.
  \bibinfo{pages}{1--6}.
\newblock
\urldef\tempurl%
\url{https://doi.org/10.1109/ASRU57964.2023.10389710}
\showDOI{\tempurl}


\bibitem[Warden(2018)]%
        {warden2018speech}
\bibfield{author}{\bibinfo{person}{Pete Warden}.}
  \bibinfo{year}{2018}\natexlab{}.
\newblock \showarticletitle{Speech Commands: A Dataset for Limited-Vocabulary
  Speech Recognition}.
\newblock \bibinfo{journal}{\emph{arXiv}}  \bibinfo{volume}{abs/1804.03209}
  (\bibinfo{year}{2018}).
\newblock


\bibitem[Zhang et~al\mbox{.}(2017)]%
        {zhang2017hello}
\bibfield{author}{\bibinfo{person}{Yundong Zhang}, \bibinfo{person}{Naveen
  Suda}, \bibinfo{person}{Liangzhen Lai}, {and} \bibinfo{person}{Vikas
  Chandra}.} \bibinfo{year}{2017}\natexlab{}.
\newblock \showarticletitle{Hello Edge: Keyword Spotting on Microcontrollers}.
\newblock \bibinfo{journal}{\emph{arXiv}}  \bibinfo{volume}{abs/1711.07128}
  (\bibinfo{year}{2017}).
\newblock


\end{thebibliography}

\end{document}